\documentclass[preprint]{emulateapj} 

\usepackage{natbib}
\usepackage{hyperref}
\hypersetup{colorlinks,citecolor=Blue,linkcolor=Red,urlcolor=Blue}
\usepackage{amsmath}
\usepackage{empheq}
\usepackage[usenames,dvipsnames]{color}
\allowdisplaybreaks 

\begin{document}

\title{The density of mid-sized Kuiper belt objects from ALMA thermal observations}

\author{Michael E. Brown} 
\affil{California Institute of Technology, Pasadena CA 91125 (U.S.A.)}

\author{Bryan J. Butler}
\affil{National Radio Astronomy Observatory, Socorro NM 87801 (U.S.A.)}

\begin{abstract}

The densities of mid-sized Kuiper belt objects are a key constraint into understanding the assembly of
objects in the outer solar system. These objects are critical for understanding the currently
unexplained transition from the smallest Kuiper belt objects with densities lower than that of water 
to the largest objects with significant rock content. Mapping this transition is made difficult by the
uncertainties in the diameters of these objects, which maps into an even larger uncertainty in volume and
thus density. The substantial collecting area of the Atacama Large Millimeter Array allows significantly
more precise measurements of thermal emission from outer solar system objects and could potentially greatly
improve the density measurements. Here we use new thermal observations of four objects with satellites 
to explore the improvements possible with millimeter data. We find that effects due to effective emissivity at
millimeter wavelengths make it difficult to use the millimeter data directly to find diameters 
and thus volumes for these bodies. In addition, we find that when
including the effects of model uncertainty, the true
uncertainties on the sizes of outer solar system objects measured with radiometry are likely larger
than those previously published.
Substantial improvement in object sizes will likely require precise occultation measurements.

\end{abstract}


\section{Introduction}

For small solar system bodies like Kuiper belt objects (KBOs) about
which little is known, density can be one of the key characteristics
which informs our understanding of bulk composition, formation, and
physical state. For most small bodies, measurement of density requires
separate accurate measurement of the mass and of the volume of the
system. While mass can often be measured to high accuracy by fitting a
Keplerian orbit to the motion of a small satellite, measurement of the
size of the body is often more difficult and less accurate.  While
stellar occultations have given extremely accurate estimates of a
limited number of larger bodies \citep{2016_Sicardy, 2011_Sicardy,
2006_Sicardy, 2013_Brown_a, 2013_Braga-Ribas}, most size estimates for
these small bodies rely on the much less accurate technique of thermal
radiometry.

In thermal radiometry, the thermal emission of a body at a known
distance from the sun is used to estimate the emitting area -- thus the
diameter -- of the body. To date, most measurements of thermal emission
have been made from the Spitzer Space Telescope \citep[at 24 and 70
$\mu$m;][]{2008_Stansberry} and with the Herschel Space Telescope
\citep[at 70-500 $\mu$m;][]{2010_Muller}.  Such
measurements have the advantage of spanning both the Wien and
Rayleigh-Jeans side of the thermal emission, which typically peaks
around $\sim$100 $\mu$m for bodies at these temperatures. 
The disadvantage of these measurements,
however, is the moderately low signal-to-noise ratio (SNR) of the
thermal emission at these wavelengths with these telescopes and also the
sensitivity of the size measurements to model assumptions such as the
thermal properties, spin period, and pole position of the objects 

The ALMA radio observatory provides the potential for thermal
observations of KBOs with a complementary set of advantages and
disadvantages. With the massive collecting area of ALMA, thermal
observations can be obtained at significantly higher SNR.  Unfortunately,
these observations are at longer wavelength and thus exclusively on the
Rayleigh-Jeans tail of the thermal emission, preventing an accurate
constraint on the temperature and temperature distribution of the KBO
surface. Interestingly, however, the Rayleigh-Jeans portion of the
emission is only linearly dependent on temperature, instead of
exponentially.  More uncertainly, there is the possibility that at these
long wavelengths emissivity-like effects depress the thermal emission in
unknown ways \citep{1998_Muller, 2013_Fornasier, 2016_Lellouch}, making
prediction of thermal flux densities (or of radii from thermal
measurements) unreliable.  We note that a similar depression has been
seen in the flux densities of asteroids at long wavelengths, but this is
not due to a true emissivity effect, but rather the properties of the
subsurfaces of these bodies \citep[i.e., when a proper radiative
transfer model is used with realistic surface and subsurface properties,
no depressed emissivity is needed to fit the observations --
][]{2013_Keihm}.  To date, with no available Spitzer/Herschel
measurements along with high signal-to-noise longer wavelength
measurements we cannot assess the effect of such measurements on our
ability to measure size -- and thus density -- of KBOs.

We explore the utility of ALMA measurements of KBO size of a carefully
selected group of objects. As noted in \citet{density_paper}, the apparent
increase in density with diameter of known KBOs is difficult to explain
in conventional accretion scenarios.  The smallest bodies have densities
below 1 g cm$^{-3}$ suggesting both significant porosity and significant
ice fraction, while the largest bodies have densities greater than 2 g
cm$^{-3}$, implying a substantial rock fraction. The difficulty of
creating the large bodies by accretion of the small bodies could be
overcome if the small bodies have similar rock fractions to the larger
bodies but are significantly more porous. The mid-sized KBOs provide the
key to answering this question. At diameters greater than about 600 km
little porosity can be sustained inside of KBOs \citep{2010_Yasui}, so
the density measured comes close to reflecting a true ice-to-rock ratio.
\citet{density_paper} found that 2002 UX25, with a diameter of 692 $\pm$
23 km, had a density of 0.82 $\pm$ 0.11 g cm$^{-3}$, suggesting that,
indeed, the smallest bodies were incapable of accreting to form the
larger bodies. Here we attempt to improve the density measurements of a
sample of 4 mid-sized KBOs in order to more carefully explore this
transition from small-to-large bodies and to determine if the results
found from 2002 UX25 carry through to this larger sample.

\section{Observations}

All observations were undertaken with the 12-m 
array of the Atacama Large Millimeter Array (ALMA).  This synthesis
array is a collection of radio antennas, each 12 m in diameter, spread
out on the Altiplano in the high northern Chilean Andes.  Each of the
pairs of antennas acts as a two element interferometer, and the
combination of all of these individual interferometers allows for the
reconstruction of the full sky brightness distribution, in both
dimensions \citep{2001_Thompson}.

ALMA is tunable in 7 discrete frequency bands, from $\sim$90 to $\sim$
950 GHz.  All observations in this paper were taken in Bands 6 and 7,
near 230 and 350 GHz, in the ``continuum'' (or ``TDM'') mode, with the
standard frequency tunings.  For band 6, this yields four spectral
windows in the frequency ranges: 220--222 GHz; 222--224 GHz; 236--238
GHz; and 238--240 GHz.  For band 7 the frequency ranges are: 337--339
GHz; 339--341 GHz; 349--351 GHz; and 351--353 GHz.  In the final data
analysis we average over the entire frequency range in both bands, and
use 230 GHz and 345 GHz as the effective frequencies in our modeling.

All of these observations are in dual-linear polarization; in the end
we combine these into a measurement of Stokes I.  While we expect 
polarized emission from the surfaces, it is weak and in an unresolved
image averages to zero.

Table \ref{obs-tab} shows the information for all of our observations;
most of which were executed in November 2013 (with a single observation
in March 2014).  The number of antennas included in the observation was
between 24 and 35, but in all but the March 2014 observation of Quaoar
at band 7 there were three antennas that were significantly distant
from the others which provided no useful signal and had to be removed 
from the processing.  The array was
mostly in the C32-3 configuration, which yields a resolution of $\sim$0.9"
in band 6 and $\sim$0.6" in band 7 - the actual resolution for each of the
observations is shown in Table \ref{obs-tab}.  Each observation was of
order 1 hour in duration, including all calibration overheads, which
resulted in roughly 20-30 minutes on source.  Neptune and Titan were
used as the absolute flux density scale calibrators for all
observations \citep{2012_Butler}.  Nearby point-like radio source
calibrators were used to calibrate the phase of the atmosphere and
antennas as a function of time.
\begin{deluxetable*}{ccccccc}
\tablecaption{ Observing dates and geometry.  \label{obs-tab}}
\tablehead{
\colhead{body} & \colhead{band} & \colhead{date/time} & \colhead{Distance} & \colhead{resolution} & \colhead{primary}    & \colhead{secondary} \\
\colhead{}      &\colhead{}      & \colhead{(UTC)}     & \colhead{(AU)}     & \colhead{(arcsec)}   & \colhead{calibrator} & \colhead{calibrator} \\
}
\startdata
2002 UX25 & 6 & 2013-Nov-18/02:00-02:45 & 40.22 & 0.81 X 0.72 & Neptune & J0238+1636 \\
2002 UX25 & 7 & 2013-Nov-05/04:10-04:35 & 40.18 & 0.60 X 0.54 & Neptune & J0231+1322 \\
Salacia   & 6 & 2013-Nov-04/23:05-24:06 & 43.80 & 0.94 X 0.73 & Neptune & J2253+1608 \\
Salacia   & 7 & 2013-Nov-05/00:30-01:10 & 43.80 & 0.58 X 0.57 & Neptune & J2253+1942 \\
Quaoar    & 6 & 2013-Nov-04/22:10-22:45 & 43.77 & 1.20 X 0.63 & Neptune & J1733-1304 \\
Quaoar    & 7 & 2014-Mar-21/08:10-08:24 & 42.95 & 0.82 X 0.45 & Titan   & J1733-1304 \\
Orcus     & 6 & 2013-Nov-04/12:40-12:55 & 48.37 & 0.83 X 0.67 & Titan   & J1007-0207 \\
Orcus     & 7 & 2013-Nov-16/12:30-12:55 & 48.19 & 0.80 X 0.68 & Titan   & J1058+0133 \\
\enddata
\end{deluxetable*}

Initial calibration of the data was provided by the ALMA observatory,
and is done in the CASA reduction package via the ALMA pipeline
\citep{2014_Muders}.  The actual measured quantity of a complex
interferometer like ALMA is a sampling of the complex visibility
function at the positions of the baselines between each of its antennas.
The visibility function is the two dimensional Fourier transform of the
sky brightness distribution.  The individual samples of the visibility
function are referred to as visibilities, and are complex quantities
(real and imaginary, or amplitude and phase).  After the initial
calibration, the data product was a set of visibilities for each of the
observing dates.

\begin{figure}
\plotone{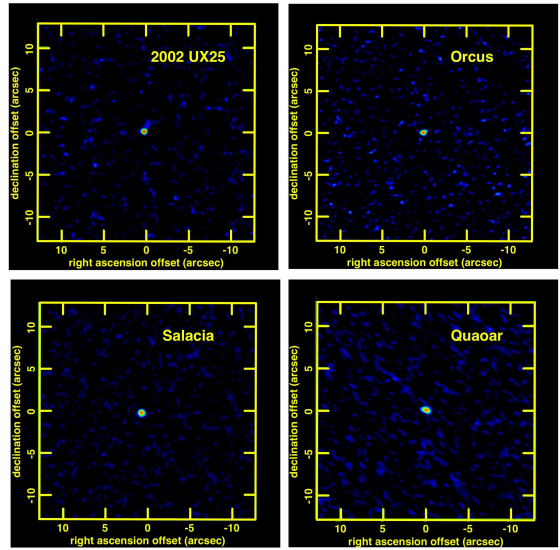}
\caption{ALMA Images of 2002 UX25, Orcus, Salacia, and Quaoar at the Band 7 frequency
of 350 GHz (870 $\mu$m). \label{band7images}}
\end{figure}

\begin{figure}
\plotone{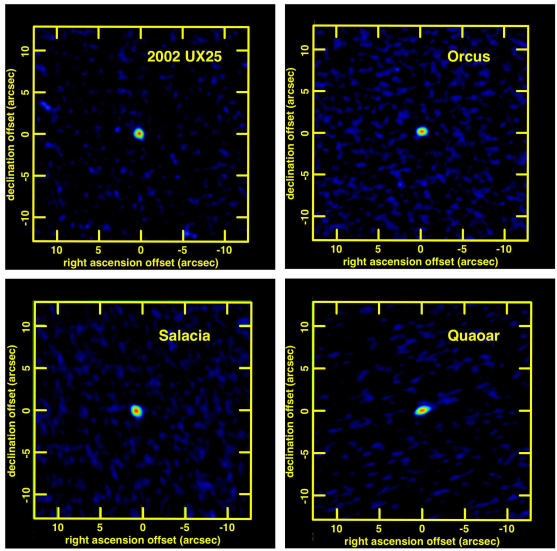}
\caption{ALMA Images of 2002 UX25, Orcus, Salacia, and Quaoar at the Band 6 frequency
of 230 GHz (1300 $\mu$m). \label{band6images}}
\end{figure}

At this point we exported the data from CASA and continued the data
reduction in the AIPS package
(www.aips.nrao.edu/CookHTML/CookBook.html).  We imaged each of the
objects at each band with natural weighting.  The result is shown in
Figures \ref{band7images} (band 7) and \ref{band6images} (band 6).  The
final step of the data analysis was to estimate the observed flux
density for each body in both bands.  We obtained this value in a number
of ways, to check for consistency: flux density in the image; flux
density in the CLEAN components; fitting a gaussian in the image; and
fitting the visibilities directly.  We found good agreement (better than
the final 1$\sigma$ uncertainty) for all of
these techniques.  We take the visibility fit value as the best value,
as it avoids the biases of fitting in the image plane in the presence of
correlated noise \citep{2004_Greisen}.  As a byproduct of the fitting
process, an offset from the phase center is also derived.  Table
\ref{data-tab} shows the final fitted flux densities and offsets for our
observations.
\begin{deluxetable}{cccc}
\tablecaption{ Derived flux densities and offsets.  \label{data-tab} }
\tablehead{
 \colhead{body} & \colhead{band} & \colhead{flux density} & \colhead{Offsets} \\
\colhead{}      &\colhead{}      & \colhead{(mJy)}        & \colhead{(RA; Dec - arcsec)} \\
}
\startdata
2002 UX25 & 6 & 0.49 $\pm$ 0.03 & .166 $\pm$ .017; -.016 $\pm$ .016 \\
2002 UX25 & 7 & 1.06 $\pm$ 0.06 & .184 $\pm$ .014; -.010 $\pm$ .014 \\
Salacia   & 6 & 0.71 $\pm$ 0.02 & .649 $\pm$ .011; -.174 $\pm$ .012 \\
Salacia   & 7 & 1.33 $\pm$ 0.05 & .637 $\pm$ .008; -.200 $\pm$ .009 \\
Quaoar    & 6 & 0.62 $\pm$ 0.02 & -.098 $\pm$ .018; -.031 $\pm$ .011 \\
Quaoar    & 7 & 1.89 $\pm$ 0.07 & -.107 $\pm$ .011; -.031 $\pm$ .008 \\
Orcus     & 6 & 0.85 $\pm$ 0.04 & -.124 $\pm$ .014; .103 $\pm$ .012 \\
Orcus     & 7 & 1.44 $\pm$ 0.12 & -.079 $\pm$ .019; .095 $\pm$ .015 \\
\enddata
\end{deluxetable}

\section{Thermal modeling}

\subsection{Background}

To first order, thermal emission from a body depends on only the
distance to the sun and the surface albedo, which together set input
heat flux per unit area, and the diameter of the body, which sets
emitting area. Multiple additional properties, however, can change true
temperature distribution on the surface of the body, leading to
different fluxes and distributions of thermal emission. These additional
properties are parametrized in various thermal emission models as
surface roughness, surface thermal inertia, or a catch-all correction term
called the ``beaming factor.'' Most models of KBO thermal emission
use some variant of a ``standard thermal model'' in which it is assumed
that all thermal emission is directed toward the sun, implying
either a pole-on configuration or zero thermal inertia. The beaming
factor is then adjusted to raise or lower the assumed surface
temperature in order to match the spectrum of the observed thermal
inertia. In such a model the beaming factor approximately accounts
for changes from the simplistic equilibrium surface temperature.
Non-zero
thermal inertia and non-polar orientations cause radiation on the non-illuminated
side of the body and thus lower the expected surface temperature.
Surface roughness
causes sunward- (and thus, for distant KBOs, observer-) facing
surface facets to receive higher
insolation and reach higher temperature. This second effect is the
beaming to which the beaming factor refers. When used to refer to
actual beaming, this factor can only have the effect of increasing
the surface temperatures.  When the beaming factor is also used as a
proxy for unknown thermal inertia and pole position, however, it can
change the temperature in either direction. 

The largest thermal emission studies of KBOs to date 
use this simple pole-on (or, equivalently, zero thermal inertia) thermal
model modified by an effective beaming factor. Limited studies of the
highest signal-to-noise data have explored the use of a more complex
thermo-physical model which explicitly models the various thermal
parameters \citep{2010_Muller, 2013_Fornasier}, but
\citet{2013_Lellouch} has shown that, for these data, the simple model
gives equivalent results. We will thus adopt the same general approach.

Our focus in this work is on both obtaining the most accurate diameter
measurements possible, but also in understanding the true uncertainties
in these diameter measurements. We are thus as interested in the
uncertainties generated by the assumed model as we are by the
observational uncertainties. An accurate assessment of both is critical
for understanding whether or not our density measurements constrain the
mode of formation of objects in the outer solar system. 


\subsection{Multi-parameter Markov chain Monte Carlo thermal modeling}

We model thermal emission from these KBOs using a standard thermal
model with a beaming factor which is adjusted to account for the combined
effects of thermal inertia, pole position, and surface roughness.
Our thermal model takes as input the radius, $R$, 
beaming factor, $\eta$,  and Bond
albedo $A$.  The surface temperature at angle $\Theta$ away from the sub-solar point
is calculated as
$$T= \Big[{{S\cos(\Theta)(1-A)}\over{\sigma\epsilon\eta}}\Big]^{1/4},$$
where $\epsilon$ is the bolometric emissivity -- discussed below -- $\sigma$ is the
Stefan-Boltzmann constant, and $S$ is the solar insolation at the
distance of the object.  
The thermal
emission from the observer-facing area is then integrated to calculate
the total emission at each wavelength.  The possible range of bolometric 
emissivity, $\epsilon$, for
KBOs, is unclear. Bolometric emissivities of Pluto and Charon have
been estimated to be between 0.83 and 0.93 \citep{2016_Lellouch}.
In 
well-measured asteroids, emissivities 
vary from about 0.8 to 0.9 \citep{1998_Muller}.
We will take values from 0.8 to 1.0 to encompass the range of possible
uncertainty and allow our emissivity to vary between these values.
Typical KBO thermal models assume a fixed emissivity of 0.9.

The model also calculates the expected absolute visible magnitude,
$H_V$, of the object from its radius and visible geometric albedo,
$p_V$, as $$H_v=-5 \log{_{10}(D p_v^{1/2} / 1330)}, $$ where $D$ is the
diameter, in kilometers.  The visible geometric albedo, $p_v$, which
determines the zero-phase reflected sunlight, is connected to the Bond
albedo, which determines the energy absorbed, through the phase
integral, $q$, as $A=q p_v$.  No phase integrals have been measured for
KBOs, as high phase observations have been unavailable, but
\citet{2009_Brucker} shows that the icy Saturnian and Uranian satellites
roughly follow a linear function given by $q=0.336p_v+0.479$.
Significant outliers occur, however, with Phoebe -- perhaps a good
analog for darker KBOs -- being a factor of two below this value, and
Europa being a factor of two above. We take this factor of two below or
above variation to represent the uncertainty in our knowledge of the
phase integral and allow our phase integral to vary by this factor from
the Brucker fit.

The models are compared to the observations of the flux density as a
function of frequency and to the measured absolute visible magnitude
using a Markov chain Monte Carlo (MCMC) model in which the free
parameters are diameter, albedo, and beaming factor,
with variations in emissivity and phase integral and the uncertainties
in measured absolute magnitude also included as
nuisance parameters.  The individual thermal observations were obtained
when the object was at different heliocentric and geocentric distances,
so, though the differences are small, each flux density is modeled for
the individual distance that the observation was made at.  We use the
Python package {\it emcee} \citep{2013_Foreman-Mackey}, which provides a
convenient and parallel implementation of the \citet{2012_Hou} affine
invariant ensemble sampler for MCMC. In each of the modeling cases that
we discuss below, we assign uniform priors to the parameters 
and run an ensemble of 100 chains
through $10^4$ steps after a $10^3$ step initialization (``burn-in'')
period.  The chains converge with no obvious memory of initial condition.
We examine marginalized distributions of all of the parameters. The
distribution of diameter is nearly gaussian, thus we report the median
and the middle 68.2\% range to represent the 1 $\sigma$ uncertainties. 

\section{Spitzer/Herschel fits}

\begin{figure}
\plotone{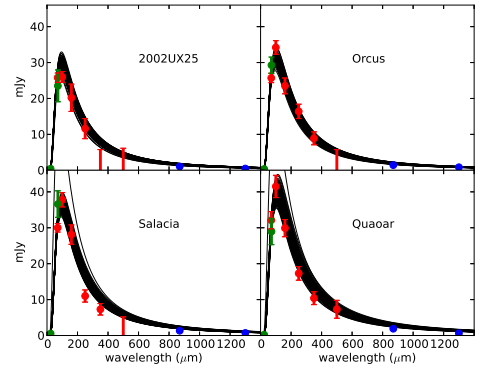}
\caption{Examples of MCMC fits using only the Spitzer (green) and Herschel (red)
thermal data shortward of 350 $\mu$m.
The ALMA fluxes
are show (blue), but not included in these model fits. The uncertainties on the
ALMA measurements are smaller than the data points. The curves are a
	collection of 100 samples from the MCMC
	ensemble which illustrates the statistical range of the acceptable fits.} 
\label{modelfit-spitzer-herschel}
\end{figure}

First, we explore the effects of the model uncertainties on the
previously estimated diameters of our four target KBOs.
\citet{2013_Fornasier} suggests that for many objects effective
emissivity drops perhaps longward of 350 $\mu$m, so we limit the initial
modeling to shorter wavelengths to explore this effect
using only the data from Spitzer and the shorter Herschel bands.  Figure
\ref{modelfit-spitzer-herschel} shows samples of the thermal model from
the Markov chain for the four targets, while Table
\ref{diameter-fit-tab} gives derived diameters and albedos. Note that
\citet{2013_Fornasier} report detections that are less than 1 $\sigma$
as upper limits at the 1 $\sigma$ flux level, which makes these fluxes
impossible to use in any statistical way; as we have no additional
knowledge of the actual measured flux level, we run a simple Monte Carlo
model to estimate that the correct 1 $\sigma$ upper limit to a randomly
distributed signal that is measured to be somewhere between 0 and 1
$\sigma$ is 1.86 $\sigma$; we accordingly use those values for the
limits, though we do not incorporate them into the fits.
\begin{deluxetable}{cccc}
\tablecaption{ 
   Equivalent diameter and albedo fit
\label{diameter-fit-tab} }
\tablehead{
 \colhead{body} & \colhead{albedo,} & \colhead{diameter (km),} & \colhead{diameter (km),}\\
 \colhead{}     & \colhead{this work} & \colhead{this work} & \colhead{previous\tablenotemark{1}}
}
\startdata
2002 UX25 & 0.10$\pm$0.01 & 698 $\pm$ 40  & 697 $\pm$ 40  \\
Orcus     & 0.23$\pm$0.02 & 965 $\pm$ 40  & 958 $\pm$ 22   \\
Salacia   & 0.042$\pm$0.004 & 914 $\pm$ 39  & 901 $\pm$ 45  \\
Quaoar    & 0.12$\pm$0.01 & 1083 $\pm$ 50 & 1073 $\pm$ 38  \\
\enddata
\tablenotetext{1}{\citet{2013_Fornasier}}
\end{deluxetable}

The results from Table \ref{diameter-fit-tab} show that while using a
more realistic estimate of model and parameter uncertainties does not
change the best fit values of the diameters and albedos, it increases
the uncertainties in some of these values. The uncertainties in the diameters
are increased, on average, by a factor of about 25\%, albeit with
considerable variation. 
While this value is not extreme, it does
lead to a 75\% increase in the uncertainty of the measurement of
the volume and thus the density of these objects.  We find that the
uncertainty increase is dominated by 
the expanded range of parameters such as the
emissivity and phase integral.  The models do not give substantially
different best-fit results, but the range of diameters over which
adequate fits can be obtained is clearly higher.

\begin{figure}
\plotone{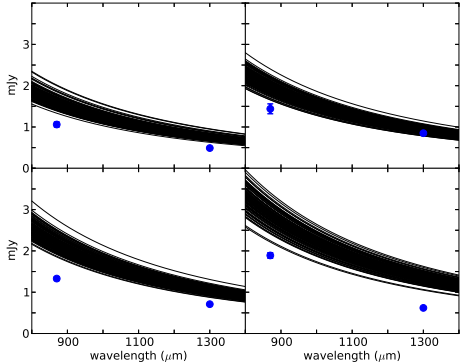}
\caption{The same model fits using only Spitzer and Herschel data shortward of
350 $\mu$m now
compared to the ALMA fluxes. The shorter wavelength data significantly over predict
the fluxes in the ALMA bands. When not visible, the uncertainties on the data are
smaller than the plotted points.}
\label{modelfit-spitzer-herschel-alma}
\end{figure}

While appropriate use of the upper limits makes it less clear that the
longest Herschel wavelengths suffer from depressed emission due to
decreasing effective emissivities, Figure
\ref{modelfit-spitzer-herschel-alma}, which shows an expanded view of
these Spitzer/Herschel-only fits to the ALMA observations, shows a clear
discrepancy. The ALMA flux densities are systematically about 30\% lower
than predicted from the Spitzer/Herschel observations.  With the high
signal-to-noise ratio of the ALMA observations, these discrepancies are
highly significant. In the next section we explore whether these higher
signal-to-noise observations can improve the estimates of the diameters.

\section{Spitzer/Herschel/ALMA fits and millimeter emissivity}

\begin{figure}
\plotone{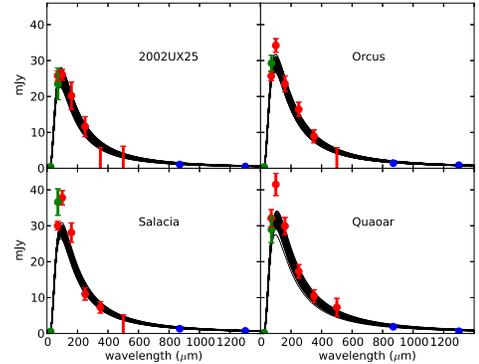}
\caption{Examples of MCMC model fits using all Spitzer, Herschel and ALMA
    data simultaneously (with the exception of the uncertain 350 and 500 $\mu$m Herschel data). 
While the models now fit the low-uncertainty 
ALMA points more closely, the discrepancies at the $\sim$100 $\mu$m 
emission peak severe.}
\label{modelfit-all}
\end{figure}

With the significantly smaller uncertainties, the ALMA measurements have
the possibility of significantly improving the diameter uncertainties.
We use our MCMC model to fit the full suite of Spitzer, Herschel, and
ALMA data (with the exception of the 350 and 500 $\mu$m Herschel data,
which we still exclude because of the uncertainty about the
uncertainties).  Figure \ref{modelfit-all} shows the fits to this full
data set.  In most cases, the small uncertainties of the ALMA results
drives the model fits to significantly under predict the Spitzer and
Herschel data near the peak of the thermal emission, while
still over predicting the ALMA flux densities.  Best fit diameters are
decreased by $\sim$15\%, and while the formal uncertainties decrease by as 
much as 25\%, the actual fits appear quite poor.

We can conceive of only two likely explanations for this effect. Either
there is a serious difference in calibration between ALMA and the
infrared space telescopes, or effective emissivity effects at longer
wavelengths are indeed suppressing the longer wavelength thermal
emission on these bodies. 

A calibration uncertainty of this magnitude in the ALMA data
is unlikely, given the care
with which these data are calibrated.  Flux densities measured by ALMA
are specified to have an accuracy of $\pm$5\%, and all tests of this
claim to date have verified this accuracy.  There have been some recent
questions regarding the use of monitored quasars and asteroids to set
the flux density scale for ALMA observations, but in our case only
Neptune and Titan were used for this, and confidence is high that the
models for these two bodies are good \citep{2012_Butler}.

\begin{figure}
\centering
\plotone{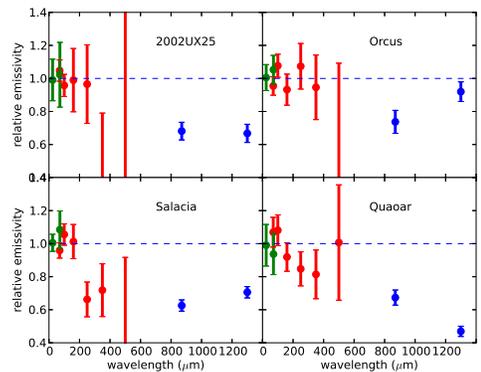}
\caption{Relative effective
emissivity as determined using only the Spitzer+Herschel model.}
    \label{emissivity}
\end{figure}

A suppressed millimeter emissivity seems the most likely cause of the
low ALMA flux densities. \citet{2013_Fornasier} discuss decreasing
emissivity at 350 and 500 $\mu$m in the Herschel data, but with the
large uncertainties at these wavelengths the true effect is difficult to
determine. With the ALMA data, the discrepancy becomes unmistakable.  In
Figure \ref{emissivity} we use the Spitzer/Herschel only fits of the
previous section to determine the effective emissivity as a function of
wavelength. 

Note that \citet{2016_Lellouch} have also found a depressed emissivity
at these long wavelengths in observations of the Pluto/Charon system.
They interpret this depression as due to a combination of surface and
subsurface dielectric and particle/volume scattering rather than from
surface roughness effects (which have often been invoked for these and
other bodies at the shorter thermal emission wavelengths, and as
described above are actually part of the genesis of the beaming
factor).  A similar depression on the four KBOs we have observed here
may be due to the same effect.

\section{Effects of satellites}

Some satellites are potentially large
enough to have a significant effect on the total thermal emission. If the
primary and the satellite have identical surface properties (and pole
positions), their emission spectra are identical and the individual
contributions simply scale as the squares of their diameters. In this
simplest case the diameter of the primary can be obtained from the
equivalent diameter, $D_e$, calculated above, and the difference in
magnitude between the primary and satellite $\Delta V$, as
$D^2=D_e^2(1+10^{-\Delta V/2.5})^{-1}$. 

KBO primaries and their satellites have highly correlated surface
colors \citep{2009_Benicchi}, leading to the reasonable expectation that the
surfaces are similar and have similar enough properties that this
simplest case prevails.  The Orcus-Vanth system clearly deviates from
this simple expectation. Orcus has significant water ice absorption that
Vanth is lacking, and they have differing optical colors \citep{2010_Brown}. The derived
system albedo of 0.23$\pm$0.02 is high for typical KBOs but not unusual for large
ones with visible water ice.  All of these considerations strongly
suggest that Vanth is darker than Orcus.  In such a case Vanth will be
warmer than Orcus and a simple scaling cannot be used to estimate their
sizes.  We run an additional MCMC model to simultaneously fit the
emission from a potentially darker Vanth. In this model we
force Orcus and Vanth to
have the same beaming factor, bolometric emissivity, and phase integral;
these assumptions could be
suspect if the two bodies have very different surfaces.
We fix Vanth to be 2.61 magnitudes fainter
than Orcus \citep{2010_Brown}, and add the constraint that the albedo of Vanth 
must be less than or equal to the albedo of Orcus. The MCMC model fits these
models to all of the data out to 250 $\mu$m, appropriately accounting for
thermal emission from,
for example, a large darker warmer Vanth, or a small bright colder Vanth. 
The median of the marginalized posterior distribution of the
diameter is 885$^{+55}_{-80}$ km for Orcus with an albedo of 0.25$^{+.05}_{-.03}$ and a
diameter of 370$^{+160}_{-70}$ km for Vanth and an albedo of 0.13$\pm$0.06.
This model fit should not be considered a detection of Vanth, but merely a statement
of the best limits to the size of Vanth with the stated assumptions and observations.
The retrieved albedo of Vanth remains higher than typical
values for objects of that size, and, indeed, a significant tail to the
distribution extends to albedos as low as 0.04 (below that the thermal
emission from such a large Vanth becomes too significant), with
corresponding larger diameters for Vanth. We will retain our formal
error bars as calculated from the MCMC model but not discount the
possibility that Vanth could still be darker and larger.

\section{Densities}

Our main interest in calculating diameters and more rigorous
uncertainties for this set of KBOs is to calculate the densities of
these mid-sized objects. One final uncertainty that we must consider is
the possibility of a difference in the density of the primary and the
secondary.  The two KBOs which have estimates for satellite densities
show that the range can be extreme: Charon has a density nearly equal to
that of Pluto \citep{2015_Stern}, while the two satellites of Haumea have densities
several time smaller than that of Haumea \citep{2009_Ragozzine}. 
We will make the
assumption that satellite densities can range from 0.5 g cm$^{-3}$ up to
the density of the primary. 
For Salacia, Quaoar, and 2002 UX25, we simply calculate the range
of satellite masses, subtract that from the system mass, and determine
the density of the primary. For Orcus, we take the Orcus and Vanth sizes
from our Markov chain and randomly assign Vanth a density between 0.5 g
cm$^{-3}$ and the density of Orcus. We arrive at an Orcus density of
 $1.65^{+0.34}_{-0.24} $ g cm$^{-3}$. 
The derived sizes and densities of all objects
are shown in Table \ref{density}.
\begin {deluxetable*}{cccccc}
\tablecaption{Derived diameters and densities \label{density}}
\tablehead{\colhead{body} & \colhead{$\Delta$ V} & \colhead{primary diameter} & \colhead{satellite diameter} & \colhead{primary density} &\colhead{occultation diameter}\\
\colhead{} & \colhead{(mag)} & \colhead{(km)} & \colhead{(km)} & \colhead{( g cm$^{-3}$)} &\colhead{(km)}
}
\startdata
2002 UX25 & 2.28$\pm$0.06 & 659$\pm$38 & 230$\pm$19 & 0.80$\pm$0.13 & \\
Orcus & 2.54$\pm$0.01& 885$^{+55}_{-80} $ & 370$^{+170}_{-70}$ & 1.65$^{+.35}_{-.24}$ & \\
Salacia & 2.37$\pm0.06$ & 866$\pm$37 & 290$\pm$21 & 1.26$\pm$0.16&  \\
Quaoar & 5.6$\pm$0.2 & 1079$\pm$50 & 82$\pm$17 & 2.13$\pm$0.29& 1110$\pm$5\tablenotemark{a} \\
\enddata
\tablenotetext{1}{\citet{2013_Braga-Ribas}}
\end{deluxetable*}

\section{Conclusion}

The high sensitivity of ALMA at millimeter wavelengths allows precise
measurements of thermal emission from objects in the outer solar system.
Unfortunately, the lack of independent knowledge of millimeter
emissivities of these objects prevents these less uncertain measurements
from decreasing the uncertainties of diameter measurements of these
objects. Indeed, when parameter uncertainties are more
carefully included, we find that the current estimates of the
uncertainties in the sizes of KBOs are approximately a factor of 25\%
too small. 

Currently, no observatories are capable of observing the $\sim$100$\mu$m
thermal peak of KBOs, so any current size or albedo measurements will
have to rely on these longer wavelength observations. To explore how
well ALMA-only measurements would recover the sizes of our four observed
objects, we assume that the effective  emissivity of each
object is the average of that measured from the other three objects. We
then rerun our thermal emission MCMC fitting only the ALMA data scaled
by this effective emissivity (but retaining the 0.8-1.0 bolometric emissivity for
the emission peak). We find diameters as follows: 2002 UX25 is estimated to be
$742\substack{+76 \\ -109}$ km, Orcus is $1075\substack{+121 \\ -156}$ km, Salacia is
$900\substack{+95 \\ -140}$ km, and Quaoar is $1057\substack{+107 \\ -168}$ km. In all
cases the ALMA-only diameters are within 1-$\sigma$ of the
Spitzer/Herschel diameters, albeit with 
uncertainties two to three times larger. 

As a second check, we derive a diameter for Charon, which has been
measured to have a flux of 7.0$\pm$0.07 mJy at 840 $\mu$m \citep{2015DPS....4721004B}.
We use our ALMA-only model with an assumed emissivity of 0.685 -- the average of the values
for our 4 KBOs -- and find a diameter 
of 1355$\pm$110 km, which is within 1.3$\sigma$ of the measured value of 1212$\pm$6 km
\citep{2015_Stern}.

The larger uncertainties in KBO diameters inferred from the analysis
here renders the original goal -- better constraining the behavior of
density versus size for these objects -- impossible.  Assuming that most
KBO diameters measured with radiometry (rather than occultations) have
similarly underestimated uncertainties does not substantially change the
interpretation of \citet{density_paper}, but it makes further progress difficult.

The densities of mid-sized KBOs remains a key constraint for
understanding the accretional history of the solar system. Because of
the moderate model and parameter uncertainties associated with diameter
measurement from thermal radiometry combined with factor-of-three
greater effect on volume and density measurements, we conclude that
this technique is unlikely to yield the precision necessary to further
constrain these densities. A high priority should be placed on obtaining
occultation measurements of the satellite-bearing mid-sized objects.

\bibliographystyle{apj}
\bibliography{bibliography}
\end{document}